\documentclass[11pt]{article}

\usepackage{graphicx}    

\newcommand{\ket}[1]{| #1\rangle}  
\newcommand{\bra}[1]{\langle #1|}  
\def\one{{\mathchoice {\rm 1\mskip-4mu l} {\rm 1\mskip-4mu l} {\rm
1\mskip-4.5mu l} {\rm 1\mskip-5mu l}}}

\begin{document}

\title{Quantum algorithms for simulated annealing}

\author{
Sergio Boixo \thanks{
Quantum A.I. Lab, Google, Venice, CA, USA.
\texttt{boixo@google.com}
} 
\and
Rolando D.\ Somma\thanks{
Theoretical Division, Los Alamos National Laboratory, Los Alamos, NM, USA.
\texttt{somma@lanl.gov}
} 
}

\date{}

\maketitle  

%

\section{Problem Definition}
This problem is concerned with the development of quantum methods
to speed up classical algorithms based on simulated annealing (SA).

SA is a well known and powerful strategy to solve discrete
combinatorial optimization problems~\cite{SimAnn}. The search space
$\Sigma=\{\sigma_0, \ldots , \sigma_{d-1}\}$ consists of $d$
configurations $\sigma_i$ and the goal is to find the (optimal)
configuration that corresponds to the global minimum of a given cost
function $E: \Sigma \rightarrow {\bf R}$.  Monte Carlo implementations
of SA generate a stochastic sequence of configurations via a sequence
of Markov processes that converges to the low-temperature
Gibbs (probability) distribution, $\pi_{\beta_m}(\Sigma) \propto \exp(- \beta_m
E(\Sigma))$.  If $\beta_m$ is sufficiently large, sampling from the Gibbs
distribution outputs an optimal configuration with large probability,
thus solving the combinatorial optimization problem.  The annealing
process depends on the choice of an annealing schedule, which consists
of a sequence of $d \times d$ stochastic matrices (transition rules)
$S(\beta_1), \ S(\beta_2), \ldots , \ S(\beta_m)$. Such matrices are determined, e.g.,
by using Metropolis-Hastings~\cite{Hastings}. The real parameters $\beta_j$ denote a sequence
of ``inverse temperatures".  The implementation complexity of SA is
given by $m$, the number of times that transition rules must be
applied to converge to the desired Gibbs distribution (within
arbitrary precision).  Commonly, the stochastic matrices are sparse
and each list of nonzero conditional probabilities and corresponding configurations,
$\{\Pr_\beta(\sigma_j|\sigma_i),j: \Pr_\beta(\sigma_j|\sigma_i)>0\}$, can be efficiently computed on input $(i,\beta)$. 
This implies an efficient Monte Carlo implementation of each Markov process.
 When a lower bound on the spectral gap
of the stochastic matrices (i.e., the difference between the two
largest eigenvalues) is known and given by $\Delta>0$, one can choose
$(\beta_{k+1}-\beta_k) \propto \Delta/ E_{\rm max}$ and $\beta_0=0$,
$\beta_m \propto \log \sqrt d$.  $E_{\rm max}$ is an upper bound on $\max_\sigma |E(\sigma)|$. The
constants of proportionality depend on the error probability
$\epsilon$, which is the probability of not finding an optimal
solution after the transition rules have been applied.  These choices
result in a complexity $m \propto E_{\rm max}\log \sqrt d /\Delta$ for
SA~\cite{SimAnn:Cost}.

Quantum computers can theoretically solve some problems, such as integer factorization,
more efficiently than classical computers~\cite{Shor}. This work 
addresses the question of whether quantum computers could also solve
combinatorial optimization problems more efficiently or not. The
answer is satisfactory in terms of   $\Delta$
  (Key Results). The complexity of a quantum algorithm is determined
by the number of  elementary steps  needed to prepare a quantum state
that allows one to sample from the Gibbs distribution after
measurement. Similar to SA, such a complexity is given by the number
of times a unitary corresponding to the stochastic matrix is used. For simplicity, 
we assume 
that the stochastic matrices are sparse and disregard
the cost of computing each list of nonzero conditional probabilities and configurations, as well as the cost of computing $E(\sigma)$.
We also assume
$d=2^n$ and the space of configurations
$\Sigma$ is represented by $n$-bit strings.   Some assumptions can be relaxed.

\subsection{Problem}
{\itshape
\textsc{INPUT}: An objective function
$E: \Sigma \rightarrow \bf R$, sparse stochastic matrices $S(\beta)$ satisfying the detailed balance condition,
a lower bound $\Delta >0$ on the spectral gap of $S(\beta)$, an error probability $\epsilon >0$.
\\
\textsc{OUTPUT}: A random configuration $\sigma_i \in \Sigma$ such that $\Pr(\sigma_i \in S_0) \ge
1- \epsilon$, where $S_0$ is the set of optimal configurations
that minimize $E$.
}
\section{Key Results}
The main result is a quantum algorithm, referred to as quantum
simulated annealing (QSA), that solves a combinatorial optimization
problem with high probability using $m_Q \propto E_{\rm max} \log \sqrt d /
\sqrt{\Delta}$ unitaries corresponding to the stochastic matrices~\cite{QSA}.
The quantum speedup is in the spectral gap, as $1/\sqrt \Delta \ll 1/\Delta$ when $\Delta \ll 1$.

Computationally hard combinatorial optimization problems are typically
manifest in a spectral gap that decreases exponentially fast in $\log d$,
the problem size. The quadratic improvement in the gap is then most significant
in hard instances. The former QSA is based on ideas and techniques from quantum walks and the
quantum Zeno effect, where the latter can be implemented by evolution randomization~\cite{phaserandom1}.
Nevertheless, recent results on ``spectral gap amplification'' allow for other quantum algorithms
that result in a similar complexity scaling~\cite{gapamp}. 
\subsection{Quantum walks for QSA}
A quantization of the classical random walk is obtained by first
defining a $d^2 \times d^2$ unitary matrix that satisfies~\cite{walk1,walk2,walk3}
\begin{equation}
X \ket{\sigma_i} \ket {{\bf 0}} = \sum_{j=0}^{d-1} \sqrt{\Pr \!_\beta(\sigma_j|\sigma_i)} \ket{\sigma_i}  \ket{\sigma_j} \; .
\end{equation}
The configuration $ \bf 0$ represents a  simple configuration, e.g., ${\bf 0} \equiv \sigma_0 =0 \ldots 0$
(the $n$-bit string), and $\Pr_\beta(\sigma_j|\sigma_i)$ are  the entries of the stochastic matrix $S(\beta)$. 
The other $d^2 \times d^2$ unitary matrices used by QSA are $P$,
the permutation (swap) operator that transforms $\ket{\sigma_i} \ket{\sigma_j} $
into $\ket{\sigma_j} \ket{\sigma_i}$,
 and $R=\one - 2 \ket {\bf 0} \bra{\bf 0}$, the reflection operator over $\ket {\bf 0}$.
 
 The quantum walk is $W=X^\dagger P X P R P X^\dagger P X R$ and the detailed balance condition implies~\cite{QSA}
 \begin{equation}
 \label{eq:desiredstate}
 W \sum_{i=0}^{d-1} \sqrt{\pi_\beta(\sigma_i)}  \ket{\sigma_i} \ket {{\bf 0}}= \sum_{i=0}^{d-1} \sqrt{\pi_\beta(\sigma_i)} \ket{\sigma_i}\ket{{\bf 0}} \; ,
 \end{equation}
where  $\pi_\beta(\sigma_i)$ are the probabilities given by the Gibbs distribution. ($X$, $X^\dagger$, and $W$ also depend on $\beta$.) 
  The goal of QSA is to prepare the corresponding eigenstate of $W$ in Eq.~\ref{eq:desiredstate}, within certain precision $\epsilon>0$, and for inverse temperature $\beta_m \propto \log d$.
  A projective quantum measurement of $\ket{\sigma_i}$ on such a state outputs an optimal solution in the set $S_0$
  with probability $\Pr(S_0) \ge 1- \epsilon$.
  
\subsection{Evolution randomization and QSA implementation}
\begin{figure}[htb]
\centering
\includegraphics[height=7cm]{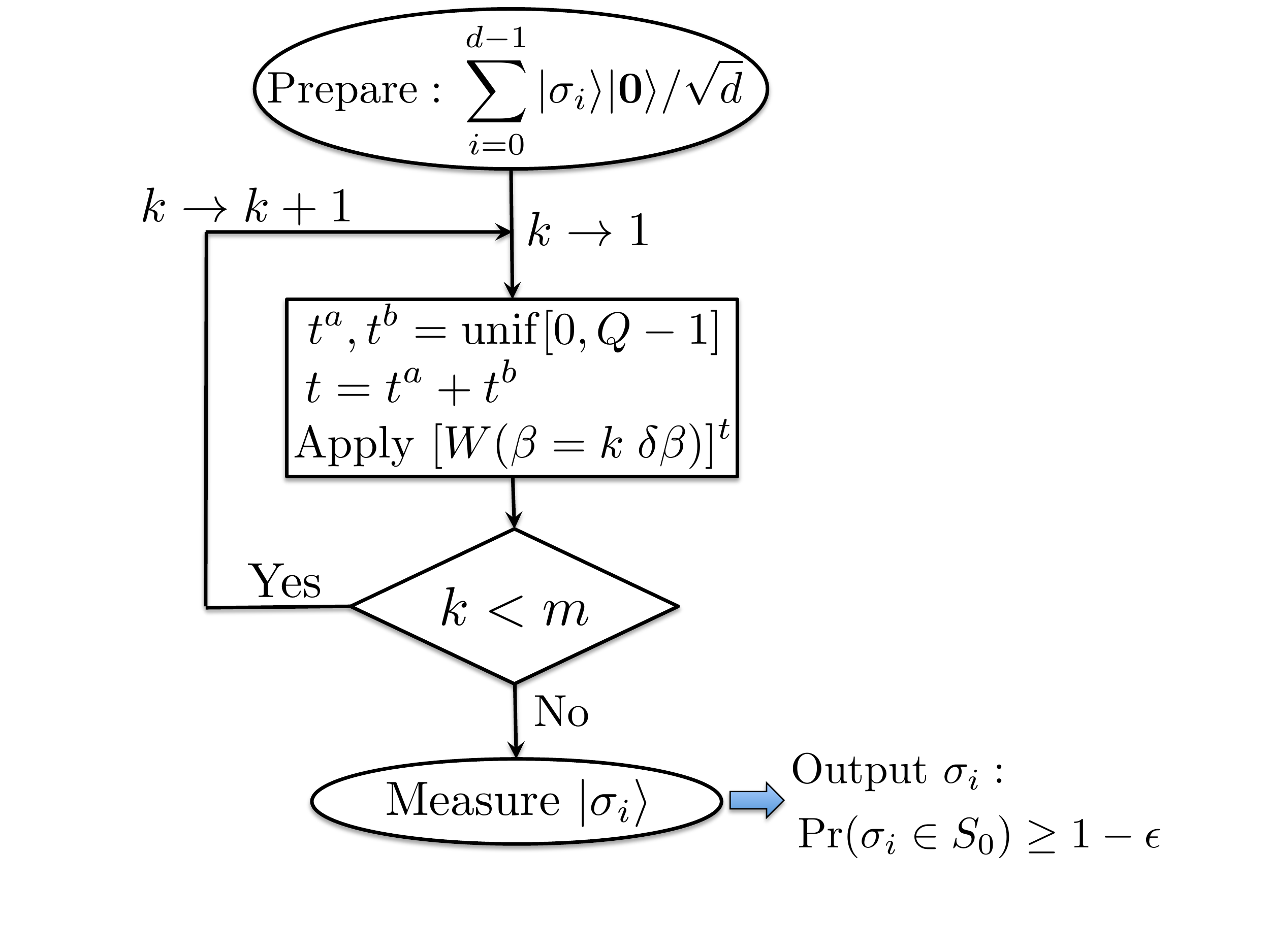}
\caption{Flow diagram for the QSA. Under the assumptions, the input state can be easily prepared on a quantum computer 
by applying a sequence of $n$ Hadamard gates on $n$ qubits. ${\rm unif}[0,Q-1]$
is the uniform distribution on nonnegative integers in that range and $Q = \lceil 2 \pi/ \sqrt\Delta \rceil$.
$\delta \beta = \beta_{k+1}-\beta_k= \epsilon/(2 E_{\rm max})$ and $m = \lceil 2 \beta_m 
E_{\rm max} /\epsilon \rceil$. Like SA, the final inverse temperature is $\beta_m = (\gamma/2) \log(2 \sqrt d/\epsilon)$,
where $\gamma$ is the gap of $E$, that is, $\gamma = \min_{\sigma \notin S_0} E(\sigma) - E(S_0)$.
The average cost of the QSA is then $mQ = \lceil 2 \pi \gamma E_{\rm max} \log(2 \sqrt d/\epsilon)/(\epsilon \sqrt \Delta) \rceil$,
and dependence on $\epsilon$ can be made fully logarithmic by repeated executions of the algorithm.
A quantum computer implementation of $W$ can be efficiently done
by using the algorithm that computes the nonzero conditional probabilities
of the stochastic matrix $S(\beta)$.}
\label{fig:1}       
\end{figure}

The QSA is based on the idea of adiabatic state transformations~\cite{phaserandom1,phaserandom2}. 
For $\beta=0$, the initial eigenstate of $W$ is $\sum_{i=0}^{d-1}   \ket{\sigma_i}  \ket {{\bf 0}}/\sqrt d$,
which can be prepared easily on a quantum computer. The purpose of QSA is then to drive
this initial state towards the eigenstate of $W$ for inverse
temperature $\beta_m$, within given precision. 
This is achieved by applying the sequence of unitary operations
$[W(\beta_m)]^{t_m} \ldots [W(\beta_2)]^{t_2} [W(\beta_1)]^{t_1}$
to the initial state (Fig.~\ref{fig:1}).
In contrast to SA, $(\beta_{k+1} -\beta_k ) \propto 1/E_{\rm max}$~\cite{phaserandom2}, but the initial and final 
inverse temperatures are also $\beta_0=0$ and $\beta_m \propto \log
\sqrt d$. This implies that the number
of different inverse temperatures in QSA is $m \propto E_{\rm max}
\log \sqrt d$, where the constant of proportionality
depends on $\epsilon$. The nonnegative integers $t_k$ can be sampled randomly according to 
several distributions~\cite{phaserandom1}. One way is to obtain $t_k$ after sampling multiple (but constant)
times from a uniform distribution on integers between $0$ and $Q-1$, where $Q = \lceil 2 \pi /\sqrt \Delta \rceil$. 
The average cost of QSA is then $m \langle t_k \rangle \propto  E_{\rm
  max} \log \sqrt d/\sqrt{\Delta}$.
One can use Markov's inequality to avoid those (improbable) instances where the cost is significantly greater than the average cost.
The QSA and the values of the constants are given in detail in Fig.~\ref{fig:1}.

\subsubsection{Analytical properties of $W$:}
The quantum walk $W$ has eigenvalues $e^{ \pm i \phi_j}$, for $j=0,\ldots, d-1$, in the relevant subspace.
In particular, $\phi_0 = 0 < \phi_1 \le \ldots \le \phi_{d-1}$ and $\phi_1 \ge \sqrt{\Delta}$~\cite{QSA,gapamp,walk1,walk2}.
This implies that the relevant spectral gap for methods based on quantum adiabatic state 
transformations is of order $\sqrt{\Delta}$. The quantum speedup follows from the fact that the complexity
of such methods, recently discussed in ~\cite{phaserandom1,phaserandom2,EPT1,EPT2},
depends on the inverse of the relevant gap.

\section{Applications}
Like SA, QSA can be applied to solve general discrete combinatorial optimization problems~\cite{COP}. 
QSA is often more efficient than exhaustive search in finding the optimal configuration.
Examples of problems where QSA can be powerful include the simulation of equilibrium states of Ising spin glasses or Potts models,
solving satisfiability problems, or solving the traveling salesman problem.

\section{Open Problems}
Some (classical) Monte Carlo implementations do not require varying an inverse
temperature and apply the same (time-independent) transition rule $S$
to converge to the Gibbs distribution. The number of times the transition rule
must be applied is the so-called mixing time, which depends on the inverse
spectral gap of $S$~\cite{mixtime}. The development of quantum algorithms to speed up
this type of Monte Carlo algorithms remains open.  Also, 
the technique of spectral gap amplification outputs a Hamiltonian 
$H(\beta)$ on input $S(\beta)$. The relevant eigenvalue of such a Hamiltonian 
is zero, and the remaining eigenvalues are $\pm \sqrt{\lambda_i}$, where   $\lambda_i \ge \Delta$. This opens the door to a quantum
adiabatic version of the QSA, in which $H(\beta)$ is changed slowly and the quantum system remains in an ``excited''
eigenstate of eigenvalue zero at all times. The speedup is also due to the increase in the eigenvalue gap. 
Nevertheless,
finding a different Hamiltonian path with the same gap,  where the adiabatic evolution occurs within   the
lowest energy eigenstates of the Hamiltonians, is an open problem. 



%
%
%
%
%
%

%
%


\end{document}